\documentclass[aps, prb, twocolumn, eqsecnum, showpacs,a4paper]{revtex4-1}
\usepackage{epsfig}
\usepackage{amsmath}
\usepackage{amssymb}
\usepackage{mathrsfs}
\usepackage{braket}
\usepackage{simplewick}
\newcommand{\be}{\begin{equation}}
\newcommand{\ee}{\end{equation}}
\newcommand{\ba}{\begin{aligned}}
\newcommand{\ea}{\end{aligned}}
\newcommand{\tr}[2]{\mathrm{Tr}_{#1}[#2]}
\newenvironment{definition}[1][Definition]{\begin{trivlist}
\item[\hskip \labelsep {\bfseries #1}]}{\end{trivlist}}

\newtheorem{theorem}{Lemma}

\begin{document}
\title{Finite-size corrections \emph{vs.} relaxation after a sudden quench}
\author{Maurizio Fagotti}
\affiliation{\mbox{The Rudolf Peierls Centre for Theoretical Physics,
    Oxford University, Oxford, OX1 3NP, United Kingdom}}
\begin{abstract}
We consider the time evolution after sudden quenches of global parameters in translational invariant Hamiltonians and study the time average expectation values and entanglement entropies in finite chains. 
We show that in noninteracting models the time average of spin correlation functions is asymptotically equal to the infinite time limit in the infinite chain, which is known to be described by a generalized Gibbs ensemble.  
The equivalence breaks down considering nonlocal operators, and we establish that this can be traced back to the existence of conservation laws common to the Hamiltonian before and after the quench. We develop a method to compute the leading finite-size correction for time average correlation functions and entanglement entropies. 
We find that large corrections are generally associated to observables with slow relaxation dynamics.
\end{abstract}

\maketitle
%%%%%%%%%%%%
\section{Introduction}%
%%%%%%%%%%%%
The experimental realization of systems that evolve unitarily for long time without dissipation~\cite{GMHB:2002,KWW:2006,HLFSS:2007,TCFMSEB:2012,CBPESFGBKK:2012} has spurred the theoretical research on nonequilibrium dynamics in isolated many body quantum systems. 
Trapped ultra-cold atom gases have become the paradigm of systems that are weakly coupled with the environment and can be designed to have highly tunable Hamiltonian parameters~\cite{B:2005}. At the same time, the evolution after a sudden quench of a Hamiltonian parameter has come to be one of the most studied protocols~\cite{SPS:2004,CC:2005,CC:2006, KLA:2007,MWNM:2007, S:2008, MK:2008, LK:2008,KE:2008,FC:2008,EKW:2009,CDEO:2008, FCC:2009,  MWNM:2009,RSMS:2009,SCC:2009,SB:2010, CEF:2011, VJSZ:2011,IR:2011,MC:2012, EEF:2012}, primarily but not exclusively because of the stationary properties that emerge at late times after the quench. 
Although the system is in a pure state, generally correlation functions as well as  reduced density matrices (RMDs) relax to stationary values\cite{LPSW:2009} which can be described in terms of an effective statistical ensemble.  
In addition, the stationary state turns out to be strongly influenced by dimensionality and local conservations laws~\cite{RDYO:2007}.

In this work we consider the time evolution of the ground state of a translational invariant spin Hamiltonian after a sudden quench of a global parameter $h_0\rightarrow h$
\begin{equation*}
\ket{\Psi(t)}=e^{-i H[h]t }\ket{\Psi_0} \qquad H[h_0]\ket{\Psi_0}=E_{GS}\ket{\Psi_0}\, .
\end{equation*}
The late time physics can be investigated in alternative ways.
In infinite chains we can study the behavior of RDMs at long time after the quench, and define the stationary state $\rho_\infty$ as follows
\be\label{eq:rhoinf}
\rho_\infty=\lim_{|A|\rightarrow\infty}\lim_{t\rightarrow \infty}\lim_{L\rightarrow\infty}\tr{\bar A}{\rho^{(L)}(t)}\, ,
\ee
where $\rho^{(L)}(t)\equiv \ket{\Psi(t)}\bra{\Psi(t)}$ is the density matrix in a chain of $L$ spins, $\bar A$ is the complement of the subsystem $A$, and we assumed the limits to exist (but there are exceptions~\cite{BS:2008}, \emph{e.g.} if there are localized excitations).

Because of quantum recurrence~\cite{BL:1957} $\rho_{\infty}$ cannot be defined in finite systems.  However generally
it turns out that at almost every time expectation values of observables are close to their time averages~\cite{GME:2011}. Thus, in a sense, the finite system relaxes to the ensemble $\bar\rho$ that describes time average expectation values
\be\label{eq:rhobar}
\ba
 &\qquad \bar \rho=\lim_{L\rightarrow\infty}\bar \rho^{(L)}\\ 
 & \tr{}{\hat O\bar \rho^{(L)}}=\lim_{T\rightarrow\infty}\frac{1}{T}\int_0^T\mathrm d t\tr{}{\hat O \rho^{(L)}(t)}\quad \forall \hat O\, .
 \ea
 \ee
It is commonly called \emph{diagonal ensemble}\cite{RDYO:2007,R:2009,R:2010,RF:2011}, since in a basis that diagonalizes the Hamiltonian only the diagonal elements of the density matrix survive the time average.

The thermodynamic limit and the infinite time limit in eqs~\eqref{eq:rhoinf}\eqref{eq:rhobar} come in different order, so a preliminary question is whether the two ensembles are identical
\be\label{eq:nonexpected}
\bar\rho\overset{?}{=}\rho_\infty\, .
\ee
In general the answer is \emph{no}. The breakdown of relations like \eqref{eq:nonexpected} in the nonequilibrium evolution of a generic pure state is a well-known fact~\cite{PSSV:2011} that can be traced back to the issue of ergodicity in quantum mechanics~\cite{N:1929}.

In this paper we focus on one aspect that rules out Eq.~\eqref{eq:nonexpected}. We indeed show that it is not unusual that the Hamiltonians before and after the quench share local conservation laws, and this is sufficient to contradict Eq.~\eqref{eq:nonexpected}.
 
A second question is whether the two ensembles have the same local properties, \emph{i.e.}
\be\label{eq:expected}
\lim_{|A|\rightarrow\infty}\tr{\bar A}{\bar \rho}\overset{?}{=}\rho_\infty\, .
\ee 
In other words we wonder whether the expectation value of local observables in finite chains are equal, on time average, to the value they approach in the limit \mbox{$1\ll t\ll L$}.
We regard Eq.~\eqref{eq:expected} as a nontrivial statement, though the physical intuition suggests \eqref{eq:expected} to be true. 
For a generic system, Eq.~\eqref{eq:expected} together with the eigenstate thermalization hypothesis~\cite{D:1991, S:1994} underlie the idea of thermal relaxation~\cite{RDO:2008}, which results in an equilibrium thermal ensemble $\rho_\infty$.
It is well known that local conservation laws after the quench invalidate this simple description, and it was proposed in Ref.~\cite{RDYO:2007} that the constraints of conservation laws $I_j$ can be taken into account by considering the generalized Gibbs ensemble (GGE)
\begin{equation}\label{eq:GGE00}
\rho^{GGE}=\frac{1}{Z}\exp(-\sum_j\lambda_j I_j)\, ,\quad\text{with}\quad [I_i,I_j]=0\, ,
\end{equation}
where $I_1=H[h]$.
 This conjecture has been verified in several systems~\cite{CC:2007, IC:2009,CE:2010, IC:2010, CEF:2011, CIC:2012} and is now expected to describe the late time behavior after a quantum quench in integrable models, in which there are infinite local conservation laws. 
Although the GGE `solves' the problem of ergodicity at infinite time after the quench for local degrees of freedom, the fact that $\rho_\infty$ can be different from $\bar \rho $ shows that there are other effects.

If Eq.~\eqref{eq:expected} is satisfied then the question moves to the finite-size corrections of $\bar \rho^{(L)}$ in Eq.~\eqref{eq:rhobar} and  
to the approach to the stationary state $\rho_{\infty}$ in the infinite chain (see also Refs~\cite{R:2010, BRI:2012,C:2006, BPGDA:2009,CCR:2011,CEF2:2012,ZS:2012}). 
In order to investigate these questions we consider quenches in models with a free-fermion representation, in which the late time behavior of correlation functions and entanglement entropies is known, and we compute the leading finite-size correction for time average correlation functions and entropies. 
We recognize that large finite-size corrections can be ascribed to the existence of  (local) conservation laws `unaffected' by the quench.
Moreover, we find that they are generally associated to observables with slow relaxation dynamics.

\subsection*{Quenches in noninteracting spin chains:\\ the price of solvability}
Noninteracting spin chains are the standard testing ground in nonequilibrium problems. If at any time the Hamiltonian can be mapped into a quadratic fermionic operator by a time independent transformation, the system can be studied by means of standard free fermion techniques. This allows numerical analyses in polynomial computational time, as well as analytic investigations.
We point up that most of the simplifications rely on the time independence of the fermion mapping.

We focus on periodic spin chains (including chains that can be interpreted as translational invariant spin ladders) and call $a_\ell$ the Majorana fermions in terms of which the Hamiltonian is quadratic. If the mapping is given by a Jordan-Wigner transformation the Majorana fermions can be defined as follows
\be\label{eq:JW}
a_{2\ell}=\prod_{j<\ell}\sigma_j^z\sigma_\ell^x\qquad a_{2\ell-1}=\prod_{j<\ell}\sigma_j^z\sigma_\ell^y
\ee
and we have $\{a_i,a_j\}=2\delta_{i j}$. Short range and translational invariance result in a Hamiltonian of the form
\be\label{eq:Hquad}
H=\frac{1}{4}\sum_{\ell,n=1}^L a_\ell\mathcal H_{\ell n} a_n
\ee
with $\mathcal H$ a block circulant matrix (apart from boundary terms), that is to say $\mathcal H_{i j}=\bar{\mathcal H}_{(i-j)\!\!\mod (L/d)}$ are \mbox{$d\times d$} matrices ($i,j=1,\dots,L/d$, in contrast to the indices in Eq.~\eqref{eq:Hquad}).   
The dimension $d$ of the blocks depends on the spin interaction. 
On the other hand the circulant structure is a consequence of translational invariance and hence is independent of the Hamiltonian parameters. 
We note that relaxing the condition of short range is not generally sufficient to spoil the structure.

It is straightforward to show that in the infinite chain there are infinite conservations laws independent of the system details. 
Indeed, if $C$ is a generic $L/d$ circulant matrix we have
\begin{multline*}
\Bigl[\frac{1}{4}\sum_{\ell,n}a_\ell\mathcal H_{\ell n} a_n,\frac{1}{4}\sum_{\ell,n}a_\ell[C\otimes \mathrm I_{d}]_{\ell n} a_n\Bigr]=\\
\frac{1}{4}\sum_{\ell,n}a_\ell[H,C\otimes \mathrm I_{d}]_{\ell n}a_n=0\, ,
\end{multline*} 
where in the last step we used that circulant matrices commute between each others. 
This shows that in short-range noninteracting Hamiltonians translational invariance produces infinite conservation laws independent of the Hamiltonian parameters.
Moreover, their range is essentially given by the size of the block of diagonals around the main diagonal that include all nonzero elements of $C$, so they can be defined to have finite range.  

The dynamics can be solved because it reduces in fact to a $d$-dimensional problem, and hence implicitly because of the infinite local integrals of motion independent of the system details. 
Although quenches in noninteracting models have been widely investigated, not much emphasis has been placed on this point and, to our knowledge, no observed behavior has been ascribed to this very aspect.  We are going to show that its signature can be found in the finite-size corrections and in the relaxation dynamics of observables.

The manuscript is organized as follows. In Section~\ref{s:model} we introduce the model and report~\cite{CEF1:2012} the explicit expressions of Eqs~\eqref{eq:rhoinf}\eqref{eq:rhobar}. In Section~\ref{s:PE} we show that the diagonal ensemble in the transverse-field Ising chain (TFIC) describes noninteracting fermions and we construct the noninteracting representation; we also discuss the effects of common conservation laws. Section~\ref{s:expvalues} is quite technical: we prove Eq.~\eqref{eq:expected} in the TFIC and develop a formalism to compute the leading finite-size correction of correlation functions. In Section~\ref{s:examples} we study the finite-size corrections for the two-point functions of the most important operators of the model. We also compute the finite-size correction for the R\'enyi entropy $S_2$ and discuss the correction for the von Neumann entropy. 

%%%%%%%%%%%%%%%%%%%
\section{The model}\label{s:model}%
%%%%%%%%%%%%%%%%%%%

In order to draw a comparison between relaxation to the stationary state in the thermodynamic limit and finite-size corrections in finite chains we focus on quenches in the transverse field Ising chain, where recently the time dependence of correlation functions has been computed exactly~\cite{CEF:2011, CEF1:2012, CEF2:2012, EEF:2012,SE:2012}. We note that our discussion generalizes straightforwardly to other spin chains with free fermion spectra, like the quantum XY model.

The Hamiltonian is given by
\be\label{eq:Ising}
H=-J\sum_{\ell}[\sigma_{\ell}^x\sigma_{\ell+1}^x+h\sigma_\ell^z]\, .
\ee
At zero temperature the model exhibits ferromagnetic ($h<1$) and paramagnetic ($h>1$) phases separated by a quantum critical point $h_c=1$. It is the simplest paradigm of quantum critical behavior and quantum phase transitions, and in the last years has also become a crucial paradigm of quench dynamics.
From a technical point of view, the Hamiltonian~\eqref{eq:Ising} with periodic boundary conditions is mapped into two separate noninteracting fermionic sectors by the Jordan-Wigner transformation~\eqref{eq:JW} and is finally diagonalized by a Bogoliubov transformation in momentum space~\cite{LSM:1961,P:1970} ($d=2$). 

In the finite chain the state before the quench, namely the ground state of the model with a given magnetic field $h_0$, is the squeezed coherent state~\cite{SFM:2012}
\be\label{eq:GS}
\ket{\Psi_0}\propto\exp\Bigl(i \sum_{0<p\in \mathrm{NS}}K(p)\alpha^\dag_{-p}\alpha^\dag_{p}\Bigr)\ket{0}\, ,
\ee
where $\ket{0}$ is the vacuum of the $\alpha$ fermions that diagonalize the antiperiodic (Neveu-Schwarz) sector of the final Hamiltonian as follows
\be\label{eq:Halpha}
H_{NS}=\sum_{p\in\mathrm{NS}}\varepsilon_{h}(p)\Bigl(\alpha^\dag_p\alpha_p-\frac{1}{2}\Bigr)\, ;
\ee
$\varepsilon_{h}(p)=2J\sqrt{1+h^2-2 h \cos p}$ is the dispersion relation and $K(p)$ is an odd function that depends on the quench details
\begin{equation*}
K(p)=\frac{(h_0-h)\sin p }{\frac{\varepsilon_{h_0}(p)\varepsilon_{h}(p)}{(2J)^2}+1+h h_0-(h+h_0)\cos p}\, .
\end{equation*}
Momenta are quantized as $p=2\pi(n+1/2)/L$, with $n$ integer.
The time evolution of $\ket{\Psi_0}$ is obtained by appending the phase $e^{i\varepsilon_p t}$ to the creation operators $\alpha^\dag_p$.

In Ref.~\cite{CEF1:2012} both $\bar\rho$ and $\rho_\infty$ have been computed.
The density matrix $\rho_\infty$ has been identified with the GGE
\begin{multline}\label{eq:GGE}
\rho_\infty\equiv\rho^{GGE}=\\
\lim_{L\rightarrow\infty}\exp\Bigl(\sum_p\log K^2(p)n_p-\log(1+K^2(p))\Bigr)\, ,
\end{multline}
where $n_p=\alpha^\dag_p \alpha_{p}$. 
We remind that local integrals of motion in Ising-like models are linear combinations of mode occupation numbers\cite{G:1982, P:1998, FE:2012} (analogous relations hold true in a quantum field theory~\cite{FM:2010}), hence ensemble~\eqref{eq:GGE} is the GGE~\eqref{eq:GGE00}, exponential of a linear combination of  local conservation laws. 
The diagonal ensemble $\bar\rho$ can be obtained easily by virtue of the factorization of the time average density matrix in the time averages of the reduced density matrices (RDMs) associated to quasiparticles with opposite momenta, \emph{i.e.}\footnote{In principle Eq.~\eqref{eq:PE} (as well as Eq.~\eqref{eq:GS}) is not correct in the thermodynamics limit after quenches from the ferromagnetic phase ($h_0<1$) because it does not take into account the spontaneous magnetization of the initial state.
However the effect disappears in the thermodynamic limit at infinite time after the quench (see also Ref.~\cite{FE:2012}).
In addition, also considering in a finite chain the evolution of the state that approaches the correct ground state of the model in the thermodynamic limit, the correction to Eq.~\eqref{eq:PE} turns out to be negligible in large chains with respect to the effects that we are discussing.  

We also point out that for special values of $h$ and $L$ there are accidental degeneracies which are not described by Eq.~\eqref{eq:PE}.}
\be\label{eq:PE}
\bar\rho\equiv\rho^{PE}=\prod_{\mathrm{NS}\ni p>0}\Bigl(\frac{1- n_p- n_{-p}}{1+K^2(p)}+n_p n_{-p}\Bigr)\, .
\ee
In Ref.~\cite{CEF1:2012} it was called \emph{pair ensemble} for emphasizing the total correlation between quasiparticles with opposite momentum
\be\label{eq:totcor}
\braket{n_k n_{-k}}_{PE}=\braket{n_k}_{PE}=\frac{K^2(k)}{1+K^2(k)}\, .
\ee
We notice instead that the GGE gives a different result (see also Ref.~\cite{GP:2008})
\be\label{eq:corGGE}
\braket{n_k n_{-k}}_{GGE}=\braket{n_k}_{GGE}^2=\Bigl(\frac{K^2(k)}{1+K^2(k)}\Bigr)^2\, .
\ee 
This is not unexpected since $n_k n_{-k}$ is a nonlocal operator and the GGE describes only local degrees of freedom~\eqref{eq:rhoinf}.
By direct comparison of Eqs~\eqref{eq:GGE}\eqref{eq:PE} (or Eqs~\eqref{eq:totcor}\eqref{eq:corGGE}) we see that \eqref{eq:nonexpected} is not satisfied after quenches in Ising-like models.

We note that in both ensembles $\rho^{GGE}$ and $\rho^{PE}$ the von Neumann entropy $S_{v.N.}[\rho]\equiv -\tr{}{\rho\log\rho}$
is extensive. The entropy density 
of the pair ensemble is given by
\be\label{eq:SPE}
\frac{S_{v. N.}[\rho^{PE}]}{L}\xrightarrow{L\gg1}\int_0^\pi \frac{\mathrm d p}{2\pi}\mathcal H\Bigl(\frac{1-K^2(p)}{1+K^2(p)}\Bigr)\, ,
\ee
with $\mathcal H(x)=-\frac{1+x}{2}\log\frac{1+x}{2}-\frac{1-x}{2}\log\frac{1-x}{2}$. On the other hand the entropy density of large subsystems (of length $\ell$) in the generalized Gibbs ensemble~\eqref{eq:GGE} is double of the pair ensemble entropy density
\be\label{eq:SGGE}
\frac{S_{v. N.}[\rho_\ell^{GGE}]}{\ell}\xrightarrow{\ell\gg 1}\int_0^\pi \frac{\mathrm d p}{\pi}\mathcal H\Bigl(\frac{1-K^2(p)}{1+K^2(p)}\Bigr)\, .
\ee
In (the appendix of) Ref.~\cite{CEF1:2012} it has been pointed out that such discrepancies must be imputed to nonlocal degrees of freedom, and the authors provided an argument for the average of local-in-space operators to be equal in both ensembles, \emph{i.e.} Eq.~\eqref{eq:expected} for quenches in the TFIC (see also Ref.~\cite{CE:2010} for quenches to noninteracting bosonic systems).

As emphasized in Ref.~\cite{G:2012}, the disagreement  between Eq.~\eqref{eq:SPE} and Eq.~\eqref{eq:SGGE} indicates that the
  time average expectation values of a substantial number of observables cannot be described through the natural finite-volume generalization of the GGE (which corresponds to remove the thermodynamic limit from Eq.~\eqref{eq:GGE} so that the sum runs over the momenta quantized in the finite volume).
 It was also suggested that this could be related to the absence of  factorization in independent uncorrelated fermion-like degrees of freedom, which in the TFIC is attributed to the strong correlations between quasiparticles of opposite momenta (\emph{cf}. Eq.~\eqref{eq:totcor}). As a matter of fact, in the next section we show that the pair ensemble is factorized in noninteracting fermions, and the disagreement \eqref{eq:nonexpected} is instead due to the existence of conservation laws independent of the quench parameter (the magnetic field).
 
%%%%%%%%%%%%%%%%%%%
\section{Pair ensemble}\label{s:PE} %
%%%%%%%%%%%%%%%%%%%

As revealed by Eqs~\eqref{eq:PE}\eqref{eq:totcor}, with the time average  the pair of quasiparticles with opposite momenta emerges as the new elemental object.
In this section we  formulate a description in terms of ``pairs''. 

First, we notice that the pair ensemble \eqref{eq:PE} describes noninteracting fermions, indeed the eigenvalues of \eqref{eq:PE} can be written as
\begin{equation*}
\lambda_{\{\sigma\}}=\prod_i\frac{1+\sigma_i \nu_i}{2}\, ,
\end{equation*}
for some $\nu_i\in[0,1]$, where $\sigma_i=\pm 1$ (we notice that half of $\nu$'s is equal to $1$).
However, because of Eq.~\eqref{eq:totcor}, the noninteracting fermions cannot be linear combinations of the fermions $\alpha$ that diagonalize the Hamiltonian as in \eqref{eq:Halpha}.

We start off with a single momentum and define two fermions, namely $g^\dag_{p}$, which we improperly call \emph{ghost}, and the \emph{pair fermion} $\tilde d^\dag_p$, as follows
\be\label{eq:g}
\ba
g^\dag_p&=(1-n_{-p})\alpha^\dag_p+n_{-p} \alpha_p\\
\tilde d^\dag_p&=(\alpha^\dag_p+\alpha_p) \alpha^\dag_{-p}\, ,
\ea
\ee
with $\{g^\dag_p,g_p\}=\{\tilde d^\dag_p,\tilde d_p\}=1$ and $\{g^\dag_p,\tilde d^\dag_p\}=\{g^\dag_p,\tilde d_p\}=0$.
The pair fermion represents to all intents and purposes the ``pair'' introduced qualitatively before, indeed its presence (absence) corresponds to the presence (absence) of both fermions $\alpha_{\pm p}^\dag$. 
We have
\be\label{eq:d1}
\ba
g^{\dag}_pg_{p}&=n_p+n_{-p}-2n_pn_{-p}\\
\tilde d^{\dag}_p \tilde d_p&=n_{-p}
\ea
\ee
so that the reduced density matrix at fixed momentum $\pm p$ is given by (\emph{cf}. Eq.~\eqref{eq:PE})
\be\label{eq:PE1}
\frac{1-n_p-n_{-p}}{1+K^2(p)}+n_p n_{-p}=g_p g^\dag_p \Bigl(\frac{1+(K^2(p)-1)\tilde d^\dag_p \tilde d_p}{1+K^2(p)}\Bigr)\, ,
\ee
which is noninteracting in $g^\dag_p$ and $\tilde d^\dag_p$.

In fact, the ghost and the pair fermion defined in \eqref{eq:d1} do not satisfy the correct anticommutation relations at different momenta: ghosts anticommute between each others but pair fermions commute both with ghosts and with other pair fermions. 
This problem can be easily solved by adding a nonlocal string to the definition of pair fermions:
\be\label{eq:dok}
d^\dag_p\equiv \!\prod_{0<k<p} (2\tilde d^\dag_k\tilde d_k-1)\!\prod_{0<k\neq p}\! (2g^{\dag}_k g_{k }-1)\tilde d_p^\dag\quad (d^\dag_p d_p=\tilde d^\dag_p\tilde d_p)\, ,
\ee
so that 
\begin{equation*}
\{d^\dag_p,d_q\}=\{g^\dag_p,g_q\}=\delta_{p q}\, ,\quad\{g^\dag_p,d^\dag_q\}=\{g^\dag_p,d_q\}=0\, .
\end{equation*}
Finally we obtain
\be\label{eq:PE2}
\rho_{PE}=\prod_{0< k}g_k g^\dag_k \exp\Bigl(\sum_{0<p}\log K^2(p)d^\dag_p d_p-\log(1+K^2(p))\Bigr)\, ,
\ee
which is a product of \emph{uncorrelated fermionic degrees of freedom}.

We also write the final Hamiltonian in terms of the new fermions
\be\label{eq:Hgd}
H_{NS}=\sum_{0<p}2\varepsilon_h(p) g_p g_p^\dag\Bigl(d^\dag_p d_p-\frac{1}{2}\Bigr)\, .
\ee
It is rather surprising that the noninteracting representation of $\bar\rho$ (\emph{cf.} \eqref{eq:PE2} and  \eqref{eq:PE}) does not correspond to the noninteracting representation of the Hamiltonian (\emph{cf.} \eqref{eq:Hgd} and \eqref{eq:Halpha}).

\subsection*{The side effect of common conservation laws}
Comparing Eq.~\eqref{eq:PE2} with Eq.~\eqref{eq:GGE} suggests that Eq.~\eqref{eq:nonexpected} is not satisfied in Ising-like models because of the ghost degrees of freedom.
They enter Eq.~\eqref{eq:PE2} as the projector on their vacuum, and in particular 
\be\label{eq:gnoPE}
\tr{}{\rho_{PE}g^\dag_p g_p}=0\, .
\ee 
In addition, $g^\dag_p g_p=(n_p-n_{-p})^2$ (\emph{cf.} Eq.~\eqref{eq:d1}) is a conserved quantity which is independent of the quench details: $n_p-n_{-p}$ is invariant under a Bogoliubov transformation of the $\alpha$ fermions, \emph{i.e.} $n_p-n_{-p}=n^{(0)}_p-n^{(0)}_{-p}$ with $n^{(0)}_p$ the occupation numbers of the Hamiltonian before the quench.
The expectation value of ghost occupation numbers is indeed zero at any time $\tr{}{\rho(t)g^\dag_k g_k}=0$ (also before the quench) and, despite the definition in Eq.~\eqref{eq:g} depends on the Hamiltonian parameters, one can even define ghosts independent of the system details; the drawback is that the Gaussian structure of pair fermions is lost, although the pair ensemble is still proportional to the projector on the ghost vacuum. 
In any case $g^\dag_p g_p$ is a conserved quantity both before and after the quench.
The GGE~\eqref{eq:GGE} does not correctly describe the expectation value of ghost occupation numbers   
\be\label{eq:gnoGGE}
\tr{}{\rho^{GGE}g^\dag_p g_p}=2\Bigl(\frac{K(p)}{1+K^2(p)}\Bigr)^2
\ee  
and since the right hand side is always positive, any linear combination of $g^\dag_p g_p$ has nonzero expectation value, in contrast to Eq.~\eqref{eq:gnoPE}: $g^\dag_p g_p$ are intrinsically nonlocal.  
This is not totally surprising if we remind that \emph{the initial and final Hamiltonians share half of the integrals of motion}\cite{G:1982,P:1998,FE:2012}. Ghosts describe the projector on the eigenspaces of the common conservation laws in which the initial state is found; they are nonlocal because generally the projector on an eigenspace of a local integral of motion is nonlocal.

Associating zero effective temperature ($\lambda\rightarrow \infty$ in Eq.~\eqref{eq:GGE00}) to the ghost modes identifies~\eqref{eq:PE2} with a  generalized Gibbs ensemble. However, it cannot be produced as a limit of reduced density matrices, as in Eq.~\eqref{eq:rhoinf}, because the limiting procedure naturally orders  the integrals of motion by their range of interaction (see also Ref.~\cite{FE:2012}) and the corresponding Lagrange multipliers do not have a proper thermodynamic limit.

Similar situations can be expected also after quenches in interacting (integrable) models. For instance, in the Heisenberg XXZ chain the interaction is invariant under a rotation, and hence the projector of the total spin on the rotation axis is a local conservation law independent of the anisotropy parameter $\Delta$ of the model. Quenching $\Delta$ results in a density matrix $\bar\rho$~\eqref{eq:rhobar} in which it is possible to factorize a projector.  
In analogy with the noninteracting case, in a basis in which the projector factorizes  as the ghost vacuum does in Eq.~\eqref{eq:PE2}, the remainder could have the form of a GGE. However, also in this case the diagonal ensemble should not be `equal' to the generalized Gibbs ensemble~\eqref{eq:rhoinf}. 
 
Another simple argument that shows that something can go wrong when there are common conservation laws is the following. 
If the state at late time after the quench is described by a GGE,  the alternative time evolution with a local Hamiltonian that commutes with the original  one (after the quench) and with all associated conservation laws should give rise to the same stationary state. 
However, choosing a common conservation law as new Hamiltonian makes the density matrix to not evolve at all, and hence the local properties to remain the same of the initial state (which, among other things, is a low-entangled state).

In the next section we  (re)derive Eq.~\eqref{eq:expected} in Ising-like models (we formalize the argument of Ref.~\cite{CEF1:2012}) and 
develop a formalism to obtain expectation values in the pair ensemble at $\mathcal O(L^{-1})$.

%%%%%%%%%%%%%%%%%%%%%%%%%
\section{Expectation values}\label{s:expvalues} %
%%%%%%%%%%%%%%%%%%%%%%%%%

The pair ensemble has a very simple form when expressed in terms of the Bogoliubov fermions~\eqref{eq:PE} but computing spin correlation functions can be cumbersome also for the simplest operators.

For any given observable, the time average can be generally replaced by the average over a finite set of variables, reducing the problem complexity;
however these approaches scale exponentially with the number of fermions that represent the observables, \emph{i.e.} with the distance in the two-point function of operators with a nonlocal fermionic representation (as the order parameter in the TFIC). 
This makes impossible to study the behavior of two-point functions at large distances and entanglement entropies of large subsystems.

In this section we show that working with ghosts and pair fermions allows us to exploit the asymptotic non-interacting nature of the density matrix and hence compute correlations at $\mathcal O(L^{-1})$ in polynomial computational time (in addition, some correlations are finally written in a form suitable for analytical investigations). 

%%%%%%%%%%%%%%%%%%%%%%%%%%%%%%%%%%%%%%%%%%%%%%%%%%%
\subsection{Wick theorem in the pair ensemble: \mbox{the removal of the ghost degrees of freedom}} %
%%%%%%%%%%%%%%%%%%%%%%%%%%%%%%%%%%%%%%%%%%%%%%%%%%%

We call \emph{local} the observables with a spin representation independent of the chain length.  In particular, spin correlation functions in the thermodynamic limit are local, regardless of the distance between operators. 
Strings of spins $\sigma_{\ell_1}^{\alpha_1}\cdots \sigma_{\ell_n}^{\alpha_n}$ are represented by strings of the Majorana fermions~\eqref{eq:JW}, hence $a_\ell^x$ and $a_\ell^y$ are the basic local objects\footnote{In fact, the Jordan-Wigner transformation~\eqref{eq:JW} is nonlocal, but nonlocality is manifested only considering disjoint subsystems~\cite{IP:2010,FC:2010,F:2012}, dynamical correlations~\cite{RSMSS:2010,FCG:2011, EEF:2012}, or in general operators made up of odd numbers of $\sigma^{x,y}$, which in our case have zero expectation values.}. 
They are linear combinations of the Bogoliubov fermions $\alpha$ that diagonalize the Hamiltonian~\eqref{eq:Halpha}
\be\label{eq:axay}
\ba
a_\ell^x&=\frac{1}{\sqrt{L}}\sum_p e^{-i p \ell}e^{i\theta_p/2}(\alpha^\dag_p+\alpha_p)\\
a_\ell^y&=\frac{1}{\sqrt{L}}\sum_p e^{-i p \ell}e^{-i\theta_p/2}i (\alpha^\dag_p-\alpha_p)\, ,
\ea
\ee
where $\theta_p$ is the Bogoliubov angle
\begin{equation*}
e^{i\theta_p}=\frac{2J (h-e^{i p})}{\varepsilon_{h}(p)}\, .
\end{equation*}
On the other hand, the inverse transformation of \eqref{eq:g}\eqref{eq:dok} is given by
\be\label{eq:inverseb}
\ba
\alpha^\dag_p&=(1-n^d_p)g^\dag_p+n^d_p g_p\\
\alpha^\dag_{-p}&=\prod_{0<k<p} (2 n^d_k-1)\, d^\dag_p\, e^{i\pi \mathcal N_g} (g^\dag_p-g_p)\\
\ea
\ee
where $n^d_p=d^\dag_p d_p$ ($n_p^g= g^\dag_p g_p$), 
\begin{equation*}
e^{i\pi \mathcal N_g}=\prod_{0<k<\pi} (2n^g_k-1)\, ,\quad  \{e^{i\pi \mathcal N_g},g_p\}=[e^{i\pi \mathcal N_g},d_p]=0\, ,
\end{equation*}
and in particular
\begin{equation*}
n_{-p}=n^d_p\qquad n_p=n_p^g+n_p^d-2 n_p^g n_p^d\, .
\end{equation*}
The standard free fermion techniques for computing expectation values rely on the linearity of Eq.~\eqref{eq:axay}, so the nonlinear mapping~\eqref{eq:inverseb} makes calculations more difficult. 
However, Eq.~\eqref{eq:inverseb} is linear in ghost operators (regarding $e^{i\pi \mathcal N_g}$ as an independent operator) and, moreover,   
\emph{ghosts commute with all other pieces}
\be\label{eq:ghostcomm}
[g^{(\dag)}_p,n_q^d]=\Bigl[g^{(\dag)}_p,\prod_{0<k<q} (2 n^d_k-1)\, d^\dag_q\, e^{i\pi \mathcal N_g}\Bigr]=0\, .
\ee
Thus, restricting ourselves to ghost degrees of freedom, we can in fact apply Wick theorem. This is the heart of 
\begin{theorem}\label{L:1}
Let $a_i=\frac{1}{\sqrt{L}}\sum_{0<p<\pi}  W_i(p) g_p+ W_i(-p) g^\dag_p$ be Majorana fermions ($W_i(-p) =W_i^\dag(p)$), linear combinations of the Bogoliubov fermions~$\alpha$, then 
\begin{multline}\label{eq:L1}
G_0 a_1\dots a_{n} G_0=\frac{1}{L^{n/2}}\!\!\!\sum_{-\pi<p_1,\dots, p_{n}<\pi}\!\!\! W_1(p_1)\cdots W_{n}(p_{n})\\
\times \mathrm{Pf}\Bigl( [\theta(p_i)\delta_{p_i, \, -p_j}]_{i<j} \Bigr)G_0\, ,
\end{multline}
where we indicated with $[M(i,j)]_{i<j}$ the skewsymmetric matrix with elements $M(i,j)$ on the upper triangular part, and $G_0\equiv\prod_{0<k<\pi}g_k g^\dag_k$ is the projector on the ghost vacuum.
\end{theorem}
The proof is as follows. 
Since $[g^\dag_p,W(q)]=0$ (\emph{cf.} Eq.~\eqref{eq:inverseb} and Eq.~\eqref{eq:ghostcomm}), the ghosts can be moved to the right preserving their order. Analogously we can move to the right the first projector on the ghost vacuum in Eq.~\eqref{eq:L1}, so that the ghost operators are sandwiched between two projectors on the ghost vacuum. We then apply the Wick theorem to the resulting factor, indeed
\begin{equation*}
G_0 g_1^{(\dag)}\cdots g_n^{(\dag)} G_0=\tr{g}{G_0 g_1^{(\dag)}\cdots g_n^{(\dag)}}G_0\, ,
\end{equation*}
where the trace is over the ghost degrees of freedom, which results in the Pfaffian of Eq.~\eqref{eq:L1}.
Now that ghosts have disappeared (we are left with the projector on their vacuum) the operator $e^{i\pi \mathcal N_g}$ (which had just a corrective algebra function) can be absorbed into the definition of pair fermions. The removal of the ghost degrees of freedom is complete.

Lemma~\ref{L:1} is the connection between the noninteracting nature of the original model and the resulting strongly correlated diagonal ensemble.  
We are going to exploit Lemma~\ref{L:1} to prove Eq.~\eqref{eq:expected} and then obtain the leading finite-size correction.
The reader not interested in the details of the derivation can jump directly to Lemma~\ref{L:3} of Section~\ref{ss:Method} for the statement of the result
or Section~\ref{s:examples} for practical applications.

%%%%%%%%%%%%%%%%%%%%%%%%%%%%%%%%%%%%%%%%%%%%%%
\subsection{Thermodynamic limit and $\mathcal O(L^{-1})$ corrections}\label{ss:Method} %
%%%%%%%%%%%%%%%%%%%%%%%%%%%%%%%%%%%%%%%%%%%%%%

If we are careful not to commute operators, we can depict Lemma~\ref{L:1} by means of Wick contractions of the Majorana fermions. For instance we have 
\begin{multline*}
\tr{}{\rho_{PE}a_1 a_2 a_3 a_4}=
\tr{}{
\tilde\rho_{PE}
\bcontraction{}{a_1}{}{a_2}
\bcontraction{a_1 a_2}{a_3}{}{a_4}
a_1 a_2 a_3 a_4
}\\
-\tr{}{\tilde\rho_{PE}
\bcontraction{}{a_1}{a_2}{a_3}
\contraction{a_1}{a_2}{a_3}{a_4}
a_1 a_2 a_3 a_4}
+
\tr{}{\tilde\rho_{PE}
\bcontraction{}{a_1}{a_2 a_3}{a_4}
\contraction{a_1}{a_2}{}{a_3}
a_1 a_2 a_3 a_4}\, ,
\end{multline*}
with
\be\label{eq:PE3}
\tilde \rho_{PE}=\prod_{0<p<\pi }\Bigl(\frac{1+(K^2(p)-1)d^\dag_p d_p}{1+K^2(p)}\Bigr)
\ee
the pair ensemble~\eqref{eq:PE2} after tracing out the ghost degrees of freedom.
The contraction means that the sum of the momenta associated to the $W$ operators~\eqref{eq:L1} is zero (and the first momentum is positive).
Indeed each term of the Pfaffian makes the momenta equal (with opposite sign) two by two. We also note that each residual sum in Eq.~\eqref{eq:L1} is  associated to a factor $1/L$. 
\begin{definition}
We call \emph{tangled} two contractions in which it is necessary to commute also the Majorana fermions of those very contractions in order to bring operators with opposite momentum close to each other (they are shown in Eq.~\eqref{eq:structures}). Otherwise we say they are \emph{untangled}.  
\end{definition}
\begin{theorem}\label{L:1a}
The sums over the momenta of tangled contractions in Eq.~\eqref{eq:L1} can be restricted to different momenta (in absolute value). 
\end{theorem}
The proof is as follows.
In general there are two kinds of tangled contractions, namely
\be\label{eq:structures}
\ba
a)& &&
\cdots
\bcontraction{}{a_i}{\cdots a_j\cdots}{a_{k}}
\contraction{a_i \cdots}{a_j}{\cdots a_{k}\cdots}{a_l}
a_i\cdots a_{j}\cdots a_{k} \cdots a_l
\cdots
\\
b)& &&
\cdots
\bcontraction{}{a_i}{\cdots a_j\cdots a_{k}\cdots }{a_{l}}
\contraction{a_i \cdots}{a_j}{\cdots }{a_{k}}
a_i\cdots a_{j}\cdots a_{k}\cdots a_{l}
\cdots
\ea
\ee 
which have opposite sign in Eq.~\eqref{eq:L1}, because they are connected by the simultaneous interchange of two different rows and corresponding columns of the matrix in the Pfaffian~\eqref{eq:L1}, operation that changes the Pfaffian sign. 
The terms in the sums~\eqref{eq:L1} over the momenta associated to the contractions~\eqref{eq:structures} that have the same absolute value simplify between each other, and hence the sums can be constrained to run over different momenta. 
Since $W$s with different momenta commute, we can reorder operators in such a way that $W$s with opposite momenta are adjacent.  
For future reference we also define the \emph{tangled group} as follows:
\begin{definition}
We call \emph{tangled group} of a contraction the set of contractions in which momenta are forced to be different (by Lemma~\ref{L:1a}) from the momentum of the contraction.
\end{definition}

In terms of ghosts and pair fermions the Majorana fermions~\eqref{eq:JW} can be written as (\emph{cf.} Eq.~\eqref{eq:axay})
\be\label{eq:Mf}
a_\ell^{x(y)}=\frac{1}{\sqrt{L}}\sum_{0<p<\pi} A^{x(y)}_\ell(p) g_p+ A^{x(y)}_\ell(-p) g^\dag_p\, .
\ee
with ($p>0$)
\be
\ba
A^{x}_\ell(p)=\cos(p\ell-\frac{\theta_p}{2})[1-i\tau_p^y]+i\sin(p \ell-\frac{\theta_p}{2})[\tau_p^z+\tau_p^x]\\
A^y_\ell(p)=\sin(p\ell+\frac{\theta_p}{2})[1+i \tau_p^y]-i\cos(p \ell+\frac{\theta_p}{2})[\tau_p^z-\tau_p^x]
\ea
\ee
and $A^{x(y)}_\ell(-p)=(A^{x(y)}_\ell)^\dag(p)$. The Pauli matrices  $\tau_p^{x,y,z}$ are given by ($[\tau_p^i,\tau_q^j]=2 i\delta_{pq}\epsilon_{i j k}\tau_p^k$)
\be\label{eq:tau}
\frac{\tau_p^x+i\tau_p^y}{2}=\prod_{0<k<p}(2n_k^d-1)d^\dag_de^{i \pi \mathcal N_g}\, ,\qquad \!\!\!\tau_p^z=2n_p^d-1\, .
\ee

We agree to order the Majorana fermions in the observables $a_{\ell_1}^{\alpha_1}\cdots a_{\ell_n}^{\alpha_n}$ as follows:
\be\label{eq:ordering}
a_{j}^y<a_{j+\ell}^y<a_{j^\prime+\ell^\prime}^x<a_{j^\prime}^x
\ee
where $j,j^\prime,\ell,\ell^\prime>0$. In this way some simplifications, which are otherwise difficult to recognize, are evident. 
The contraction of adjacent Majorana fermions reads as
\be\label{eq:fg0}
 \bcontraction{}{a_{r+\ell}^x}{}{a_{r}^x}
a_{r+\ell}^x a_r^x
=\hat f_\ell\, ,\quad
 \bcontraction{}{a_{r+\ell}^y}{}{a_{r}^y}
a_{r+\ell}^y a_r^y
=\hat f_{-\ell}\, ,\quad
\bcontraction{i}{a_{r+\ell}^y}{}{a_{r}^x}
i a_{r+\ell}^y a_r^x
=\hat g_\ell\, ,
\ee
with
\be\label{eq:fg}
\hat f_\ell=\frac{2}{L}\sideset{}{'}\sum_{0<p}e^{i\ell p\tau_p^x}\qquad \hat g_\ell=\frac{2}{L}\sideset{}{'}\sum_{0<p}\tau_p^z e^{i(\ell p+\theta_p)\tau_p^x}\, ,
\ee
where the apostrophe in the sums is used to remind that the momentum is different from any other momentum in the tangled group.
We also write the contraction (which now means equality of momenta) between $\hat f$ and $\hat g$ operators:
\be\label{eq:contraction}
\bcontraction{}{\hat f_\ell}{}{\hat f_{\ell^\prime}}
\hat f_\ell\hat f_{\ell^\prime}
=\frac{2}{L}\hat f_{\ell+\ell^\prime}\, ,\quad 
\bcontraction{}{\hat f_\ell}{}{\hat g_{\ell^\prime}}
\hat f_\ell\hat g_{\ell^\prime}
=\frac{2}{L}\hat g_{\ell^\prime-\ell}\, ,\quad
\bcontraction{}{\hat g_\ell}{}{\hat g_{\ell^\prime}}
\hat g_\ell\hat g_{\ell^\prime}
=\frac{2}{L}\hat f_{\ell^\prime-\ell}\, ,
\ee
where the tangled groups of the resulting $f$ and $g$ operators are the union of the tangled groups of the contracted operators.

We now prove Eq.~\eqref{eq:expected}. 
The operator norm of $\hat f$ and $\hat g$ \eqref{eq:fg} is less then or equal to $1$. 
On the other hand, each contraction gives a contribution $\mathcal O(L^{-1})$ (\emph{cf.} Eq.~\eqref{eq:contraction}). 
Spin correlation functions are expectation values of the product of a finite number of Majorana fermions, hence the number of contractions that come out of Eq.~\eqref{eq:L1} is independent of $L$.  
At the leading order $\mathcal O(L^0)$ we can neglect such contributions and assume that all Majorana fermions have distinct momenta (we say that they have ``maximal tangled group''). The average of products of $\hat f$ and $\hat g$ factorizes in the product of the averages, in which sums run over distinct momenta:
\be\label{eq:Wickpair}
\ba
\tr{}{\tilde \rho_{PE}\hat f_{\ell}}=&\frac{2}{L}\sideset{}{'}\sum_{0<p}\cos(\ell p)\\
\tr{}{\tilde \rho_{PE}\hat g_{\ell}}=&\frac{2}{L}\sideset{}{'}\sum_{0<p}\cos(\ell p+\theta_p)\frac{1-K^2(p)}{1+K^2(p)}\, ;
\ea
\ee
however in the thermodynamic limit the constraints disappear
\be\label{eq:Wickpair1}
\ba
\tr{}{\tilde \rho_{PE}\hat f_{\ell}}\rightarrow& \delta_{\ell 0}\\
\tr{}{\tilde \rho_{PE}\hat g_{\ell}}\rightarrow&\int_{-\pi}^\pi\frac{\mathrm d p}{2 \pi}e^{i\ell p+i\theta_p}\frac{1-K^2(p)}{1+K^2(p)}\, ,
\ea
\ee
since the number of sums and terms of the Wick expansion \eqref{eq:L1} is finite and independent of $L$.
Eq.~\eqref{eq:L1} and Eq.~\eqref{eq:Wickpair1} are nothing but the Wick decomposition in the GGE~\eqref{eq:GGE}, that is to say \emph{Eq.~\eqref{eq:expected} for noninteracting Ising-like models}; but this is not the end of the story.
\emph{Eq.~\eqref{eq:Wickpair} is in fact sufficient to characterize the $\mathcal O(L^{-1})$ correction too} (see Appendix~\ref{a:1}) and, in addition,
the terms contributing at $\mathcal O(L^{-1})$ have no more than two $\hat f$'s.

We now rewrite Eq.~\eqref{eq:Wickpair} in a more practical way. 
The terms of the Wick expansion without $\hat f$ operators are trivially described by the correlators~\eqref{eq:Wickpair}, also without the constraints on the momenta, because with the ordering~\eqref{eq:ordering}  $\hat g$ operators are tangled (see Appendix~\ref{a:1}). On the other hand, the terms with two $\hat f$'s have the following form 
\begin{equation*}
\braket{\hat f_\ell\hat f_{\ell^\prime}\cdots}\sim \frac{4}{L^2}\sum_{0<k, k^\prime \atop k\neq k^\prime}\cos (\ell k)\cos(\ell^\prime k^\prime)\braket{\cdots}_{p_i\neq k,k^\prime}\, ,
\end{equation*}
where we had to exclude the momenta associated to the $\hat f$'s in the remaining expectation value (of $\hat g$ operators). 
At the first order, the two momenta in the expectation value are decoupled
\begin{equation*}
\braket{\cdots}_{p_i\neq k,k^\prime}=\braket{\cdots}+\bigl(Q(k)+Q(k^\prime)\bigr)/L+\mathcal O(L^{-2})
\end{equation*}
where $Q$ is some bounded function, which could be written explicitly by series expanding the Pfaffian in $\braket{\cdots}$.
In fact we don't need the exact expression of $Q$, because the leading correction to $\braket{\cdots}$ is $\mathcal O(L^{-1})$, but  the sum over the other momentum is $\mathcal O(L^{-1})$ as well, resulting in $\mathcal O(L^{-2})$ contribution.
Finally we obtain ($\ell,\ell^\prime>0$)
\be\label{eq:reduction0}
\braket{\hat f_\ell\hat f_{\ell^\prime}\cdots}\sim
-\frac{4}{L^2}\!\sum_{0<k}\cos (\ell k)\cos(\ell^\prime k)\braket{\cdots}=
-\frac{2\delta_{\ell \ell^\prime}\braket{\cdots}}{L}.
\ee
As a matter of fact, it is convenient to represent the Kronecker delta as a sum, like in the central expression of Eq.~\eqref{eq:reduction0}.  However, there is no need to sum over $L/2$ elements, indeed we can write
\be\label{eq:reduction}
\braket{\hat f_\ell\hat f_{\ell^\prime}\cdots}\sim-\frac{4}{L \tilde L}\sum_{0<\tilde k}\cos (\ell \tilde k)\cos(\ell^\prime \tilde k)\braket{\cdots}\, ,
\ee
for any $\tilde L>\ell+\ell^\prime$ and the sum is over $\tilde L/2$ momenta $\tilde k=2\pi (n+1/2)/\tilde L$, with $n$ integer. 
Considering local observables, the contractions of $a^x$ and $a^y$ fermions give rise to operators $\hat f_\ell$ in which the maximal $\ell$, let us say $\ell_{M}$, is independent of $L$; it is sufficient to choose \mbox{$\tilde L>2\ell_{M}$} to have a representation \eqref{eq:reduction} valid for all terms of the Wick expansion~\eqref{eq:L1} and for any chain length $L$. In this way we can easily isolate a factor $1/L$ from the expressions.

We note that each term of the (averaged) sum in Eq.~\eqref{eq:reduction} is a term of the Wick expansion of a noninteracting model with modified correlators, and taking into account the algebra of operators (Eq.~\eqref{eq:reduction} is valid only for operators ordered as in Eq.~\eqref{eq:ordering}) we finally obtain:
\begin{theorem}\label{L:3}
At $\mathcal O(L^{-1})$ the expectation value of local observables in the pair ensemble can be written as the average of the expectation values computed in $\tilde L/2$ Gaussian states:
\be\label{eq:L3}
\braket{a_1\cdots a_n}_{PE}=\frac{2}{\tilde L}\sum_{0<\tilde k}\braket{a_1\cdots a_n}_{\tilde k}+o(L^{-1})\, ,
\ee
where ($\mathrm{sgn}(0)\equiv 0$)
\be\label{eq:L3corr}
\ba
\braket{a_{r+\ell}^x a_r^x}_{\tilde k}
=\braket{a_r^y a_{r+\ell}^y}_{\tilde k}=
\delta_{\ell 0}+i\sqrt{\frac{2}{L}} \mathrm{sgn}(\ell)\cos(\tilde k \ell)\, ,\\
i \braket{a_{r+\ell}^y a_r^x}_{\tilde k}
=\int_{-\pi}^\pi\frac{\mathrm d p}{2 \pi}e^{i\ell p+i\theta_p}\frac{1-K^2(p)}{1+K^2(p)}
\ea
\ee
and $\tilde L$ is larger than the maximal difference between the indices of the $a^{x(y)}$ fermions in Eq.~\eqref{eq:L3}.
\end{theorem} 

The leading finite-size correction $\mathcal O(L^{-1})$ can be obtained by series expanding at the second order the Pfaffian associated to the expectation values at fixed momentum $\tilde k$, as we are going to do  for the longitudinal two-point function and the R\'enyi entropy.

%%%%%%%%%%%%%%%%%%%%%%%%%%
\section{Finite-size corrections}\label{s:examples} %
%%%%%%%%%%%%%%%%%%%%%%%%%%

We consider first some operators with a local fermionic representation.

Observables that are quadratic in the Majorana fermions~\eqref{eq:JW} 
\begin{equation*}
\sigma_\ell^z,\quad \sigma_\ell^x\sigma_{\ell+1}^x,\quad \sigma_\ell^y\sigma_{\ell+1}^y, \quad \text{etc.}
\end{equation*}
have \emph{exponentially small finite-size corrections}, which come out of the substitution of sums with integrals in Eq.~\eqref{eq:Wickpair}, being $e^{i \theta_p}$ and $K^2(p)$ smooth periodic functions (at least for quenches between noncritical models). These are the same corrections that arise considering the finite volume generalization of Eq.~\eqref{eq:GGE}, so they are always present but independent of the time average.
\emph{If there were no common conservation laws, we would expect only this type of corrections}; therefore, we understand any additional correction as the effect of conservation laws independent of the quench parameter.

We now consider operators consisting of four Majorana fermions. In this simple case we can compute the expectation value explicitly.
We obtain
\begin{multline}\label{eq:2pointfer}
\braket{a_{j}^y a_{j+\ell}^y a_{j^\prime+\ell^\prime}^x a_{j^\prime}^x}_{PE}\approx \\
\braket{a_{j}^y a_{j+\ell}^y a_{j^\prime+\ell^\prime}^x a_{j^\prime}^x}_{GGE}-\frac{2}{L}\delta_{\ell \ell^\prime}\, .
\end{multline}
Ignoring the exponentially small correction discussed above the finite-size correction is nonzero only if $\ell=\ell^\prime$, \emph{i.e.} for the two-point function of the operators $\sigma_1^z$, $\sigma_1^x\sigma_2^x$, and $\sigma_1^y\sigma_2^y$. 

%%%%%%%%%%%%%%%%%%
\subsection{Transverse correlation}%
%%%%%%%%%%%%%%%%%%

The two-point function of the transverse field is the expectation value of four fermions:
\be
C^{z}(\ell)\equiv \braket{\sigma_j^z\sigma_{j+\ell}^z}=-\braket{a_j^y a_{j+\ell}^ya_{j+\ell}^x a_{j}^x}\, ,
\ee
hence from Eq.~\eqref{eq:2pointfer} we have
\be\label{eq:2pointz}
C^{z}_{PE}(\ell)-C^{z}_{GGE}(\ell)\approx 2/L\, .
\ee
The difference between the 2-point transverse correlations in the GGE and in the pair ensemble is independent of the distance. 
It is worth to compare this simple result with  the approach to the stationary state. 
We remind that in the GGE the connected two-point function of the transverse field decays exponentially with the distance~\cite{CEF2:2012}.
In Ref.~\cite{CEF2:2012} it has been observed that in order to extract the correlation length $\xi_z$ of the transverse field one has to wait a time exponentially larger than the distance. In finite systems, to avoid quantum revivals, the time must be smaller than the chain length $L$ and hence $\log L\gg \ell$. 
With the time average we obtain an analogous behavior.
Indeed, because of Eq.~\eqref{eq:2pointz}, the exponential decay can be `seen' only if $e^{-\ell/\xi_z}\gg \frac{1}{L}$, that is to say
\be\ell\ll \log L\, .
\ee

%%%%%%%%%%%%%%%%%%%%%%%%
\subsection{Order parameter 2-point function} %
%%%%%%%%%%%%%%%%%%%%%%%%

\begin{figure}[tbp]
\begin{center}
\includegraphics[width=0.48\textwidth]{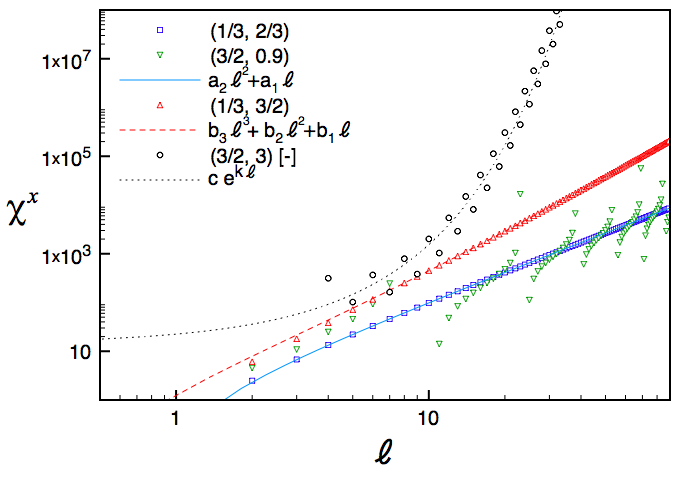}
\caption{The coefficient of the leading finite-size correction for the time average longitudinal two-point correlation for various quenches $(h_0, h)$ in the TFIC. 
In quenches from the ordered phase $\chi^x_\ell$ displays a clean power law behavior, whereas in quenches within the disordered phase (the black circles  represent $-\chi_\ell^x$) it grows exponentially with the distance. Quenches across the critical point from the disordered phase (green down-pointing triangles) are characterized by oscillations, however we recognize a parabolic growth for almost all lengths. 
}
\label{fig:Cx}
\end{center}
\end{figure}

In this section we consider the longitudinal  two point function 
\be
C^{x}(\ell)\equiv\braket{\sigma_j^x\sigma_{j+\ell}^x}\, .
\ee
In contrast to the transverse correlation the leading finite-size correction does depend on the distance, \emph{e.g.}
\be
\ba
\braket{\sigma_1^x\sigma_2^x}_{PE}&\approx \braket{\sigma_1^x\sigma_2^x}_{GGE}\\
\braket{\sigma_1^x\sigma_3^x}_{PE}&\approx \braket{\sigma_1^x\sigma_3^x}_{GGE}+\frac{2}{L}\, .
\ea
\ee
Computing $C^x(\ell)$ is much more complicated than computing $C^z(\ell)$, since the longitudinal correlation is nonlocal in terms of the Jordan-Wigner fermions. We use Lemma~\ref{L:3} to compute the leading correction. 
In each Gaussian state of eqs~\eqref{eq:L3}\eqref{eq:L3corr} the two point function has the pfaffian representation~\cite{BM:1971}
\be
\braket{\sigma_j^x\sigma_{j+\ell}^x}_{\tilde k}=\mathrm{Pf}\begin{pmatrix}
F^{(\tilde k)}&G\\
-G^t & F^{(\tilde k)}
\end{pmatrix}\, ,
\ee
with
\be\label{eq:F}
\ba
G_{m n}&=\gamma_{n-n-1}\equiv \int_{-\pi}^\pi\frac{\mathrm d p}{2 \pi}e^{i(m-n-1) p+i\theta_p}\frac{1-K^2(p)}{1+K^2(p)}\\
F_{m n}&=\phi_{m-n}\equiv i\sqrt{\frac{2}{L}}\mathrm{sgn}(m-n)\cos(\tilde k (m-n))\\
&\qquad m,n=1,\dots,\ell\, .
\ea
\ee
The function 
\be
\chi^x(\ell)=-\lim_{L\rightarrow\infty} L^2\frac{\partial}{\partial L}\log C^x_{PE}(\ell)
\ee
characterizes the leading  correction to $C^x_{PE}$
\be
C^x_{PE}(\ell)=C^x_{GGE}(\ell)\Bigl(1+\frac{\chi^x(\ell)}{L}+\mathcal O(L^{-2})\Bigr)\, .
\ee
It can be computed by series expanding the Pfaffians in Eq.~\eqref{eq:L3} at the second order in $L^{-1/2}$ and averaging over the auxiliary momentum. We finally obtain
\begin{multline}
\chi^x(\ell)
= \frac{2 L}{\tilde L}\sum_{0<\tilde k}\tr{}{(G^t)^{-1}F^{(\tilde k)} G^{-1}F^{(\tilde k)}}=\\
\sum_{i,j,k,l=1}^\ell\Bigl(\delta_{i+l\, ,k+j}-\delta_{i+j\, ,k+l}\Bigr)[G^{-1}]_{i j}[G^{-1}]_{k l}\, .
\end{multline}
In Fig.~\ref{fig:Cx} we report the numerical data for quenches within and between the ferromagnetic and paramagnetic phases of the TFIC. 
The leading correction can be neglected in the following regimes:
\be\label{eq:regimes}
\ell\ll\begin{cases}
L^{1/2}&h<1\\
L^{1/3}&h_0<1,h>1\\
\log L&h_0,h>1\, .
\end{cases}
\ee
In Refs~\cite{CEF1:2012,CEF2:2012} it has been shown that after quenches originating in the ferromagnetic phase the stationary value of $C^x(\ell)$ emerges after a time that scales as a power law with the distance $\ell$; on the other hand the timescale after which the stationary behavior reveals itself after quenches within the paramagnetic phase is exponentially large.
The finite size corrections~\eqref{eq:regimes} display the same behavior, providing further indications that \emph{large finite-size corrections in finite chains and slow relaxation in the thermodynamic limit could have the same root}, \emph{i.e.} in this specific case the existence of common conservation laws before and after the quench.
%

%%%%%%%%%%%%%%%%%%%
\subsection{R\'enyi entropy $S_2$} %
%%%%%%%%%%%%%%%%%%%

\begin{figure}[tbp]
\begin{center}
\includegraphics[width=0.48\textwidth]{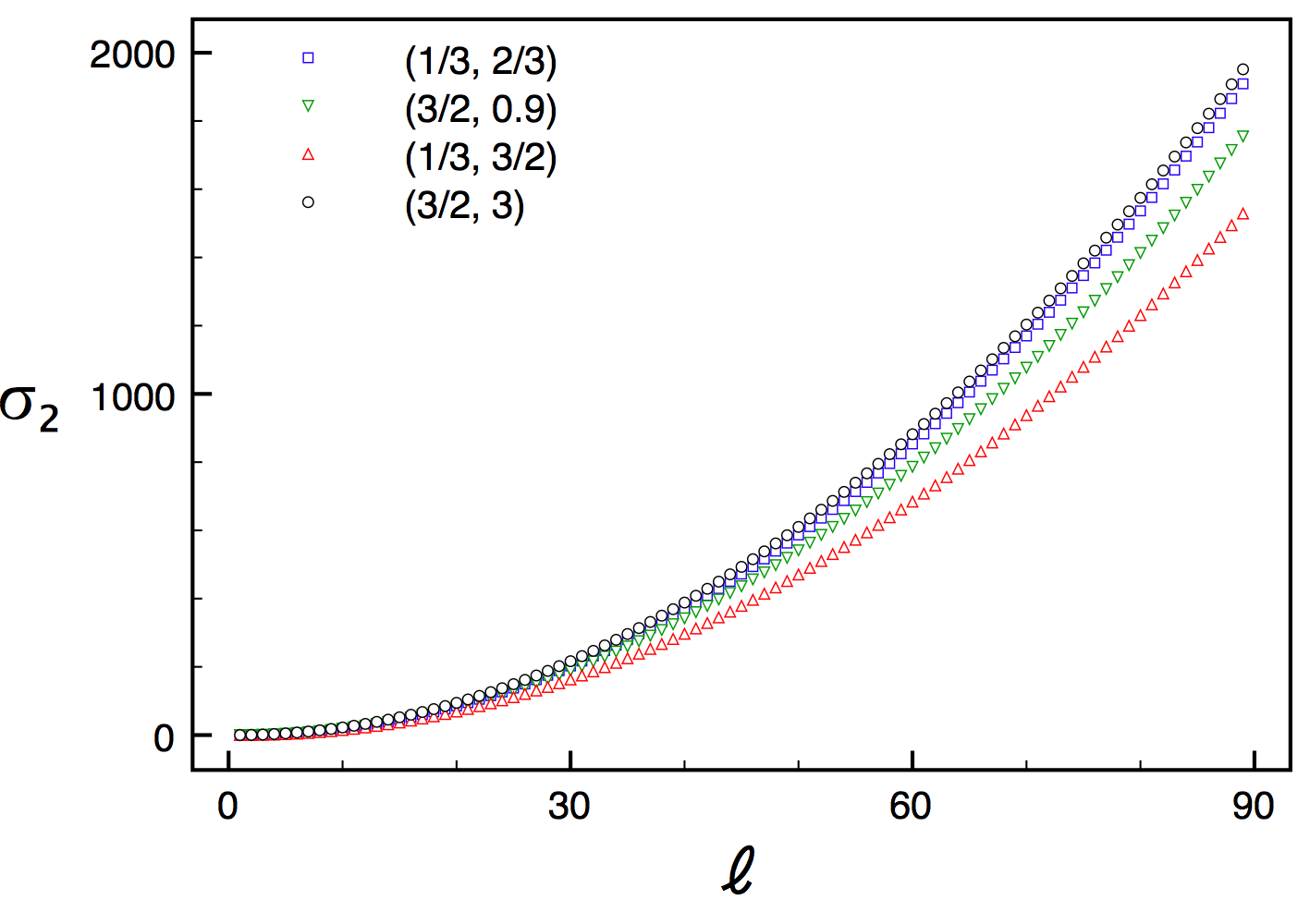}
\caption{The coefficient of the leading finite-size correction for the time average R\'enyi entropy $S_2$ for various quenches $(h_0, h)$ in the TFIC. 
In all cases the leading correction $\sigma_2(\ell)\sim \ell^2$. 
}
\label{fig:S2}
\end{center}
\end{figure}

In this section we compute the leading finite-size correction for the R\'enyi entropy $S_2\equiv-\log\tr{}{\rho_\ell^2}$, where $\rho_\ell$ is the RDM of $\ell $ adjacent spins.
We use that at the lowest order in $L^{-1}$ the RDM can be written as (\emph{cf.} Lemma~\ref{L:3})
\be
\rho_{\ell}^{PE}\sim\frac{2}{\tilde L}\sum_{0<\tilde k}\rho[\Gamma^{(\tilde k)}]\, ,
\ee
where $\rho[\Gamma^{(\tilde k)}]$ is the Gaussian matrix with correlations~\eqref{eq:L3corr}. In particular, $\Gamma^{(\tilde k)}$ is the block Toeplitz matrix
\be
\Gamma^{(\tilde k)}_{i j}=\begin{pmatrix}
\phi_{i-j}&i\gamma_{i-j}\\
-i\gamma_{j-i}&-\phi_{i-j}
\end{pmatrix}\qquad i,j=1,\dots,\ell\, ,
\ee
with $\gamma$ and $\phi$ of  Eq.~\eqref{eq:F}. We need to compute the second moment of the reduced density matrix $\tr{}{(\rho^{PE}_\ell)^2}$ 
\be\label{eq:S20}
\tr{}{(\rho^{PE}_\ell)^2}\sim \frac{4}{\tilde L^2}\sum_{0<\tilde k,\tilde k^\prime}\tr{}{\rho[\Gamma^{(\tilde k)}]\rho[\Gamma^{(\tilde k^\prime)}]}\, .
\ee
The trace of the product of two Gaussian density matrices is given by~\cite{FC:2010}
\be
\tr{}{\rho[\Gamma]\rho[\tilde \Gamma]}=\Bigl(\det\Bigl|\frac{1+\Gamma \tilde\Gamma}{2}\Bigr|\Bigr)^{1/2}
\ee
so the leading finite-size correction can be computed by series expanding the determinants in Eq.~\eqref{eq:S20} treating $\phi$'s as perturbations. 
The function 
\be
\sigma_2(\ell)\equiv \lim_{L\rightarrow \infty}L^2\frac{\partial S_2[\rho_\ell^{PE}]}{\partial L}
\ee
characterizes the leading  correction to $S_2^{PE}$
\be
S_2^{PE}(\ell)=S_2^{GGE}(\ell)-\frac{\sigma_2(\ell)}{L}+\mathcal O(L^{-2})\, .
\ee
After some algebra we obtain
\begin{multline}
\sigma_2(\ell)=
\frac{2L}{\tilde L}\sum_{0<\tilde k}\tr{}{\Bigl(\frac{1}{R^{-1}+R^t}\Bigr)^t F^{(\tilde k)} \frac{1}{R^{-1}+R^t} F^{(\tilde k)}}=\\
\sum_{i,j,k,l}^\ell\Bigl(\delta_{i+l\, ,k+j}-\delta_{i+j\, ,k+l}\Bigr)\Bigl[\frac{1}{R^{-1}+R^t}\Bigr]_{i j}[\frac{1}{R^{-1}+R^t}\Bigr]_{k l}
\end{multline}
with $F$ the matrix defined in Eq.~\eqref{eq:F} and $R_{i j}=\gamma_{i-j}$.
In Fig.~\ref{fig:S2} we report the numerical data for quenches within and between the ferromagnetic and paramagnetic phases. 
The leading correction can be neglected when $\ell\ll L^{1/2}$, but in comparison with the extensive part of the R\'enyi entropy we have 
\be\label{eq:S2cond}
\ell\ll L\, .
\ee 
Eq.~\eqref{eq:S2cond} is the time average analogous of the large time behavior of the R\'enyi entropy~\cite{FC:2008} $S_2(\ell, t)\sim S_2(\ell,\infty)+\mathcal O(\ell^4/t^3)$, where the correction can be neglected only if $\ell \ll J t$.
We point out that the leading finite-size correction is negative, as one could expect having in mind the entropy of the whole chain (\emph{cf}. eqs~\eqref{eq:SPE}\eqref{eq:SGGE} for the von Neumann entropy).  
%

%%%%%%%%%%%%%%%%%%%
\subsection{Von Neumann entropy} %
%%%%%%%%%%%%%%%%%%%

In this section we show that the correction for the von Neumann entropy is $\mathcal O(L^{-2})$.
We indicate with $\delta\rho_\ell$  the difference between the RDMs in the pair ensemble and in the GGE
\be
\rho_{\ell}^{(PE)}=\rho_{\ell}^{(GGE)}+\delta\rho_\ell\, .
\ee
At the first order in $\delta\rho_\ell$ (which corresponds to the first order in $1/L$ at fixed $\ell$  and large $L$) we have
\be\label{eq:sp1}
S_{vN}[\rho_\ell^{(PE)}]=S_{vN}[\rho_\ell^{(GGE)}]-\tr{}{\log\rho^{GGE}_\ell\, \delta \rho_\ell}+\mathcal O(L^{-2})\, .
\ee
The RDM in the GGE is the exponential of a two-form, so the trace in Eq.~\eqref{eq:sp1} is nonzero only if $\delta\rho_\ell$ has some quadratic contribution. 
This is however impossible, because the expectation value of fermionic quadratic operators is the same in the two ensembles. Thus we obtain
\be
S_{vN}^{PE}(\ell)=S_{vN}^{GGE}(\ell)+\mathcal O(L^{-2})\, .
\ee

There are indications that this result can be partially generalized to generic quenches in which Eq.~\eqref{eq:expected} is satisfied and the local properties of the system at late times after the quench are described by a GGE.

The argument is the following. 
At the first order we can extract from $\delta\rho_\ell\equiv \bar\rho_\ell-\rho^{GGE}_\ell$ the dependence on the system size as a multiplicative factor (which in Ising-like models is $1/L$) and then take the limit of infinite subsystem in the infinite system. In this limit $\rho_\ell^{GGE}$ is substituted with $\rho^{GGE}$, which is the exponential of the local conservation laws. However, by construction, the two ensembles share the same local properties, so they must give the same expectation values of local integrals of motion $I_j$, \emph{i.e.} (\emph{cf}. Eq.~\eqref{eq:GGE00})
\be
\ba
&\Delta S_{vN}=\tr{}{\log\rho^{GGE}\, \delta \rho}+\mathcal O(\delta\rho_\ell^2)\qquad \\
&\tr{}{\log\rho^{GGE}\, \delta \rho}=\tr{}{\bigl(\sum_j\lambda_j I_j\bigr) (\rho^{GGE}-\bar \rho)}= 0\, .
\ea
\ee
Thus we expect that either the first correction drops off to zero for large lengths or it is exactly zero at any length, as in  the TFIC.
However this argument can be used only to exclude the leading power law correction in $L$ and, \emph{e.g.}, does not apply when all (local) observables have exponentially small finite-size corrections.

%%%%%%%%%%%%
\section{Conclusions}%
%%%%%%%%%%%%

In this work we have shown that quenches of global parameters in periodic noninteracting chains are generally characterized by (infinite) conservation laws common to the Hamiltonians before and after the quench.
We provided evidence that they have an impact on the finite-size corrections for time average expectation values.  
Considering in particular quenches in the quantum Ising model, we identified the regimes in which the finite-size corrections are negligible and found that large corrections correspond to slow relaxation dynamics in the infinite chain~\cite{CEF2:2012}. 
This indicates that in some cases slow relaxation could be related to the fact that the initial state is (close to be) eigenstate of local conservation laws of the final Hamiltonian.

In noninteracting models the slowing down could be amplified by the fact that there are infinite common conservation laws, so this issue need further investigations. To this aim, \emph{e.g.}, it could be worth analyzing quenches in which % the (translational invariant) 
%Hamiltonian has less symmetries, for example relaxing the condition of short-range interaction. 
the noninteracting representation of the Hamiltonian depends on the quench parameter (in contrast to the Jordan-Wigner transformation~\eqref{eq:JW}). 

Finally, we did not calculated either the time average two-point functions or  R\'enyi entropies when the typical length is order the chain length.
Computing the (full) finite size scaling  appears to be a very hard problem.
In fact, semiclassical theories turned out to  reproduce the asymptotic behavior of entanglement entropies and correlation functions after a quantum quench~\cite{RSMS:2009,RSMSS:2010,RI:2011,BRI:2012,E:2012}; since there are sizable finite-size corrections for the time average entanglement entropy density (\emph{cf.} Eqs~\eqref{eq:SPE}\eqref{eq:SGGE}) we believe that a semiclassical approach might be successful also in this context.

%%%%%%%%%%%%%%
\begin{acknowledgments} 
I thank Pasquale Calabrese and Fabian Essler for useful comments. I also thank Dmitry Kovrizhin for discussions.
\end{acknowledgments} 
%%%%%%%%%%%%%%

\appendix

%%%%%%%%%%%
\section{}\label{a:1}%
%%%%%%%%%%%

Here we prove that the factorization~\eqref{eq:Wickpair} holds true at $\mathcal O(L^{-1})$.
First we observe that in each term of the Wick expansion of the operators ordered as in \eqref{eq:ordering} all $\hat g$'s are tangled, indeed untangled $\hat g$'s appear only if $a^y$ fermions come both before and after $a^x$ fermions, condition never satisfied if Majorana fermions are ordered as in \eqref{eq:ordering}. Thus, any contraction involves $\hat f$ operators. 
In addition, from Eq.~\eqref{eq:contraction} we see that the leading correction $\mathcal O(L^{-1})$ involves \emph{no more than one contraction}.
We also note that the observables with nonzero expectation values have an \emph{even number of $\hat f$}.
This is a trivial consequence of the fact that, whatever contractions are done, the contribution from $\hat f$ and $\hat g$ operators~\eqref{eq:fg} is always real. The adjoint of a string of $2n$ distinct Majorana fermions is the same string multiplied by $(-1)^{n(2n-1)}$; on the other hand each $\hat g$ operator is associated with a $i$ (\emph{cf.} Eq.~\eqref{eq:fg0}), and hence the expectation value takes a factor $(-1)^{n-n_f}$, where $n_f$ is the number of $\hat f$ operators. In order to have a nonzero expectation value  the two factors have to be equal, \emph{i.e.} $n_f$ must be even, as required.

Now we show that in the terms contributing at $\mathcal O(L^{-1})$ \emph{no more than two $\hat f$'s are present}. 

First we estimate the contribution from $\hat f$'s with maximal tangled group.
This can be inferred from the  relation (\emph{cf.} Eq.~\eqref{eq:Wickpair})
\be\label{eq:relation}
\{\ell\}_n\equiv \Bigl(\frac{2}{L}\Bigr)^{n}\!\!\!\!\!\!\sum_{0<p_1,\dots,p_{2n}\atop p_i\neq p_j,k_1,\dots k_m}\!\!\!\!\!\!\cos(\ell_i p_i)\lesssim \mathcal O(L^{-\lfloor(n+1)/2\rfloor})\, ,
\ee
valid for any $n\ll L$, $m\ll L$, $\ell_i>0$ and momenta $k_i$. 
Eq.~\eqref{eq:relation} can be proved writing the recurrence equation
\begin{multline*}
\{\ell\}_n=\delta_{\ell_n 0}\bigl(1+\frac{2}{L}(1-n-m)\Bigr)\{\ell\}_{n-1}-\frac{1-\delta_{\ell_n 0}}{L}\times\\
\Bigl[2\sum_{j=1}^m\cos(\ell_j k_j)\{\ell\}_{n-1}+\sum_{j=1}^{n-1}\sum_{\sigma=\pm 1}\{\ell\}_{n-1}^{(\ell_j\rightarrow\ell_j+\sigma  \ell_n)}\Bigr]
\end{multline*}
and noticing that we don't get a factor $\frac{1}{L}$ only when the Kronecker delta is $1$; since $\ell_j\neq 0$ the delta can be $1$  only for the terms like the last one, where lengths are shifted. In the worst case we can sum over $\lfloor(n+1)/2\rfloor$ momenta without satisfying the delta, ending up with a term in which all remaining deltas are equal to $1$. Thus, we have Eq.~\eqref{eq:relation} and in particular we see that both a single $\hat f$ ($n=1$) and two $\hat f$'s ($n=2$) with maximal tangled group contribute at $\mathcal{O}(L^{-1})$, whereas more than two $\hat f$'s give subsleading contributions.

On the other hand, if $\hat f$'s have not maximal tangled group they are contracted. 
Each contraction gives $\mathcal O(L^{-1})$ contribution (at best, \emph{cf.} Eq.~\eqref{eq:contraction}), hence we conclude that, in any case, more then two $\hat f$'s  correspond to subleading terms.

Since $\hat g$'s are tangled, terms without $\hat f$'s are simply described by Eq.~\eqref{eq:Wickpair}. We focus on terms with two $\hat f$'s.
The case of one $\hat f$ contracted with a $\hat g$ and one $\hat f$ with a maximal tangled group is subleading, since both operations are $\mathcal O(L^{-1})$. 
The contraction of the $\hat f$ operators is subleading as well because the ordering~\eqref{eq:ordering} makes positive the index of $\hat f$'s, forbidding the appearance of $\hat f_0$ as a result of the contraction (\emph{cf}. Eq.~\eqref{eq:contraction}), which is the only term that would give contribution.

 In conclusion, the tangled group of both $\hat f$'s must be maximal, so \emph{Eq.~\eqref{eq:Wickpair} is in fact sufficient to characterize the $\mathcal O(L^{-1})$ correction too}.
 
%%%%%%%%%%%%
\bibliography{ref_PE}%
%%%%%%%%%%%%

\end{document}